\documentclass[aps,preprint]{revtex4}
\usepackage{epsfig}
\usepackage{amsmath}
\usepackage{epsfig}

\begin{document}
\title
{Bound state eigenfunctions need to vanish faster than $|x|^{-3/2}$}    
\author{Zafar Ahmed}
\affiliation{Nuclear Physics Division, Bhabha Atomic Research Centre, Mumbai 400 085, India}
\email{1:zahmed@barc.gov.in}
\date{\today}
\begin{abstract}
In quantum mechanics students are taught to practice that eigenfunction of a physical bound state must be continuous and vanishing asymptotically so that it is normalizable in $x\in (-\infty, \infty)$. Here we caution  that such states may also give rise to infinite uncertainty in position $(\Delta x=\infty)$, whereas $\Delta p$ remains finite. Such states may be called loosely bound and spatially extended states that may be avoided by an additional condition that the eigenfunction vanishes asymptotically faster than $|x|^{-3/2}$.  
\end{abstract}
\maketitle
The bound state  is an essential topic in the curriculum of introductory quantum mechanics [1,2]. Bound state  eigenfunctions $\psi_n(x)$ are solutions of the Schr{\"o}dinger (1926) equation [1,2] 
\begin{equation}
\frac{d^2\psi}{dx^2}+\frac{2m}{\hbar^2}[E-V(x)]\psi(x)=0, 
\end{equation}
for some potential $V(x)$ for specific allowed values of energies $E=E_n$ that  obey Dirichlet boundary condition 
\begin{equation}
\quad \psi_n(\pm \infty)=0.
\end{equation}
Energies $E_n$ are called eigenvalues leading to quantization of energy.
This quantization was first hypothesized by Planck(1900) for the explanation of the  black body radiation and justified much later by Bohr and Sommerfeld (1919) [1,2] semi-classically for $V(x)=\frac{1}{2}m\omega^2x^2$ and other potential profiles with a single minimum called potential wells.

As $|\psi(x)|^2$ plays the role of quantal probability, $\psi(x)$ is taught to be continuous function on $x \in (-\infty, \infty)$ so that $\psi(x)$ is square integrable (normalizable) 
\begin{equation}
\int_{-\infty}^{\infty} |\psi(x)|^2 dx=1.
\end{equation}
The additional condition of differentiability of $\psi(x)$ at each and every point of the domain ensures the continuity of the momentum:
$-i\hbar \frac{d \psi(x)}{dx}$ unless there is a Dirac delta function in the potential [2]. When the potential contains a Dirac delta function e.g, $g\delta(x-a)$, the eigenfunction is allowed to become non-differentiable at $x=a$ wherein the left and right derivatives are finite but unequal at $x=a$ [3].

Further the uncertainty relation of Heisenberg (1927) [1,2]
\begin{equation}
U=\Delta x \Delta p \ge \frac{\hbar}{2}, \quad \Delta A = \sqrt{<A^2>-<A>^2}, \quad <A>=\frac{\int_{-\infty}^{\infty} \psi^*(x)~ A ~\psi(x) dx}{\int_{-\infty}^{\infty} \psi^*(x) \psi(x) dx},
\end{equation}
requires the uncertainties $\Delta x$ and $\Delta p$ to be finite for physically relevant cases. An interesting collection of uncertainty products for ground states of several solvable models may be seen in Ref.[4]. It may be mentioned that nowadays, the terms “uncertainty relation” and “uncertainty principle” are classified clearly. The term “uncertainty principle” is used when an inequality expresses the relationship between measurement error and disturbance. For a recent and a very interesting discussion on this, one may see Ref. [5].

A definite integral $\int_{a}^{b} f(x) dx$ is called proper if it  represents area under the curve $y=f(x)$ from $x=a$ to $x=b$. This in turn requires $f(x)$ to be continuous or piece-wise continuous in the closed interval $[a,b]$. On the other hand, an improper [5] integral may not connect well to the area under the curve e.g., $\int_{0}^{1} x^{-1/2} dx = 2$.  One way an integral becomes improper is when one or both of its limits are infinite. Improper integrals may or may not be convergent (finite). It is trivial yet important to see that the integral $\int_{1}^{\infty} x^{-\beta} dx$ is convergent only if $\beta>1$. This last point is crucial in determining the asymptotic behavior of $\psi(x)$ such that $<x^2>$ is finite, see below.  

The finiteness of $\Delta x$ requires $<x^2>$  to be finite, thus we demand
\begin{equation}
\int_{-\infty}^{\infty} x^2 |\psi(x)|^2 dx < \infty \Rightarrow |\psi(x)|<|x|^{-3/2}, \quad |x|\sim \infty. 
\end{equation}

On the other hand for $<p>$ and $<p^2>$ to be finite, $|\psi(x)| \sim 1/x^\epsilon, \epsilon>0$ would suffice well. Thus, the condition (5) would ensure finiteness of the uncertainty product $U$ [5].
In this paper, we point out this additional condition (5)  on the eigenfunction for a bound state.

One can write (1) in an interesting form as
\begin{equation}
V(x)-E_0=\frac{\hbar^2}{2m}\frac{1}{\psi_0(x)}\frac{d^2\psi_0(x)}{dx^2}.
\end{equation}
\noindent
Here we take $2m=1=\hbar$ which  gives $\frac{2m}{\hbar^2}=1$. It may be noted that for particle of mass 4 times that of an electron, the  approximate value of $\frac{2m}{\hbar^2}$ is $1 (eV A^0~^2)^{-1}$
when the mass and energy are in $eV$ and length is in $A^0$.
If we put the well known ground state of harmonic oscillator $\psi_0(x)=e^{-x^2/2}$ in (6), we get $V(x)-E_0=-1+x^2$. One may interpret that $V(x)=x^2$ and $E_0=1$ (or $V(x)=x^2-1$ and $E_0=0$). This $V(x)$ is the harmonic oscillator potential (See Fig. 1(a)) $V(x)=\frac{1}{2}m\omega^2 x^2$, where  $\hbar\omega = 2 eV$ and $E_0=1 eV.$ The ground state $\psi_0(x)=e^{-x^2/2}$ vanishes faster than $|x|^{-3/2}$ because as $x\rightarrow \infty$, $|x|^{3/2} e^{-x^2/2} \rightarrow 0 <1$. So the  uncertainty product $U_0$ for this ground state turns out to be finite as $U_0=\hbar/2$ [1,2], which equals the least bound  on $U$ (4) for any physical eigenstate.

Particularly, in this special case of Harmonic oscillator constructing  the raising and lowering operators [1,2] one can find all eigenvalues and corresponding eigenfunctions of $V(x)$. This is a rare feature of a potential. 

But if we put $\psi_0(x)=e^{-x^4}$ in (6), we get $V(x)=16x^6-12x^2$ and $E_0=0$. See fig. 1(b) this $V(x)$ is a double well and $E_0=0$ is the ground state of this potential that possesses infinitely many bound states. One can not construct the raising and lowering operators to obtain the rest of the spectrum of this double well potential.  If required the other bound states are obtained by numerically [7] solving the Schr{\"o}dinger equation for this potential. 
\begin{figure}[ht]
\centering
\includegraphics[width=5 cm,height=5 cm]{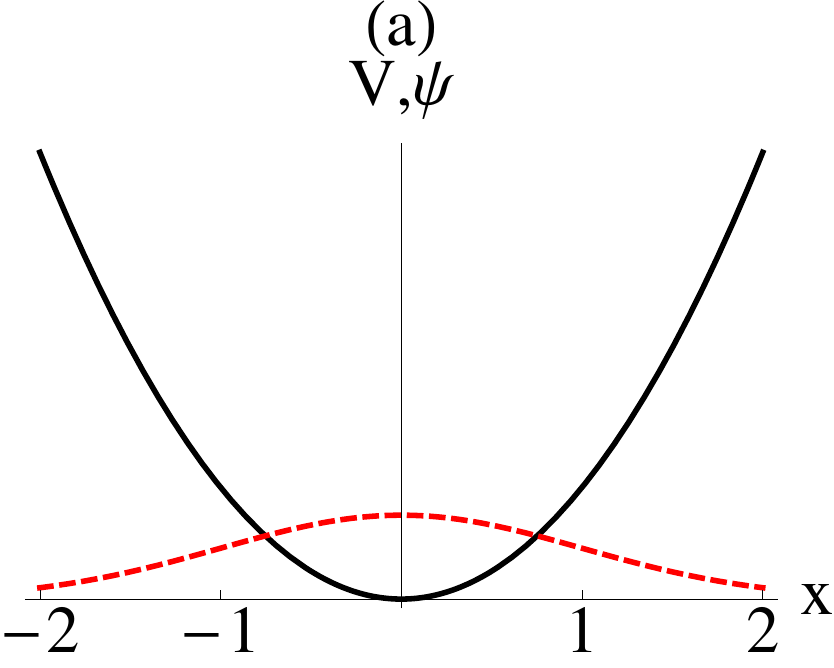}
\hskip .5 cm
\includegraphics[width=5 cm,height=5 cm]{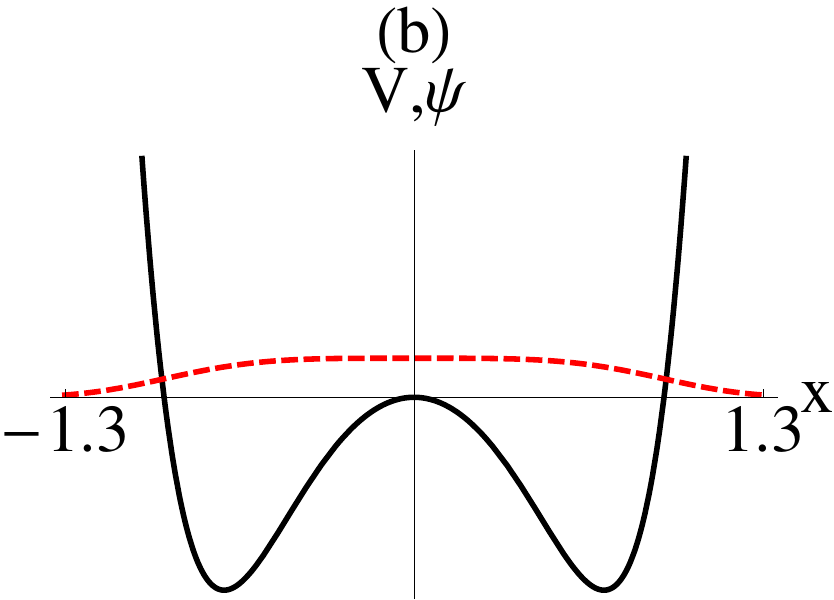}
\hskip .5 cm
\includegraphics[width=5 cm,height=5 cm]{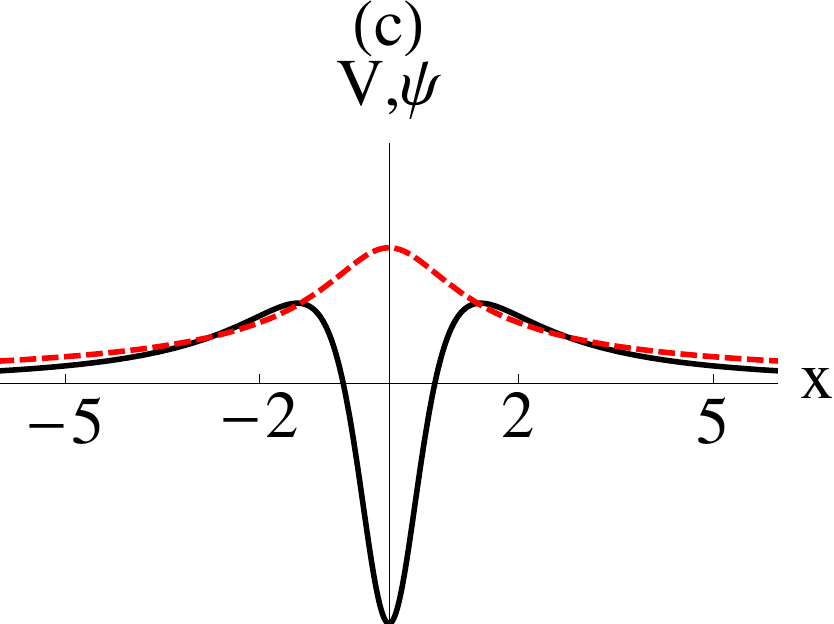}
\caption{Depiction of the normalized ground states $\psi_0(x)$ (dashed) and the corresponding potentials $V(x)$ (solid) : (a): $\psi_0(x)=(\pi)^{-1/4}~ e^{-x^2/2}$, (b): $\psi_0(x)=[2^{3/4} \Gamma(5/4)]^{-1/2}~ e^{-x^4}$, (c): $\psi_0(x)=1/\sqrt{\pi(1+x^2)}.$ Notice that in the last case: part(c), $V(x)~(16)$ vanishes as $ \sim 1/x^2$ and its ground state $\psi_0(x)$ (10) vanishes as $\sim 1/x$ [much less rapidly as compared to the previous two ground states (a,b)], hence it represents an extended state.}
\end{figure}

Once again $\psi_0(x)=e^{-x^4}$ vanishes asymptotically much rapidly as $x\rightarrow \infty$, $|x|^{3/2} e^{-x^4} \rightarrow 0 <1$ 
rendering the integrals for $<x>$ and $<x^2>$ as finite. Next, $<x>$ vanishes as there is an odd integrand within the symmetric limit of integration. Similarly, $<p>=0$ in general for bound states as they do not carry any momentum. We can find $<x^2>$
\begin{equation}
<\psi_0(x)|x^2|\psi_0(x)>= \frac{ \int_{-\infty}^{\infty} x^2 e^{-2 x^4} dx}{\int_{-\infty}^{\infty} e^{-2 x^4} dx}=\frac{\Gamma(3/4)}{4\sqrt{2}\Gamma(5/4)}
\end{equation}
and
\begin{equation}
<\psi_0(x)|p^2|\psi_0(x)>=<p\psi_0(x)|p\psi_0(x)>=16 \hbar^2 \frac{\int_{-\infty}^{\infty} x^6 e^{-2x^4} dx}{\int_{-\infty}^{\infty} e^{-2x^4} dx}=\frac{\sqrt{2} \Gamma(7/4)}{\Gamma(5/4)}.
\end{equation}
In Eqs. (7,8), we have converted the integrals to the standard Gamma functions as $\Gamma(z)=\int_{0}^{\infty} t^{z-1}e^{-t} dt $ [8].
Using (7,8) in (4), we get the uncertainty product for ground state as
\begin{equation}
U_0=\hbar \frac{\sqrt{\Gamma(3/4)\Gamma(7/4)}}{2\Gamma(5/4)}=0.5854 ~\hbar,
\end{equation}
which is more than $\hbar/2$ as per the uncertainty principle.

Similarly, the ground state $\psi_0(x)=\sqrt{\frac{2}{\pi}}\frac{1}{1+x^2}$ vanishes a little faster than $|x|^{-3/2}$ to have a well known value for $U_0=\hbar/\sqrt{2}$ [3].

Now let us take a continuous, differentiable and normalized state as
\begin{equation}
\psi_0(x)=\frac{1}{\sqrt{\pi(1+x^2)}},
\end{equation}
which behaves as $1/x$ as $x \sim \infty$. For this state
\begin{equation}
<x>=\frac{1}{\pi} \int_{-\infty}^{\infty} \frac{x}{1+x^2} dx, \quad <x^2>=\frac{1}{\pi} \int_{-\infty}^{\infty} \frac{x^2}{1+x^2} dx,
\end{equation}
so both of these integrals (11) are improper. Under the condition
that the integration range becomes infinite while maintaining the
left-right symmetry of the integration range,  we get
\begin{equation}
\Delta x=\infty.
\end{equation}
On the other hand the integral in
\begin{equation}
<p>=\frac{i\hbar}{\pi} \int_{-\infty}^{\infty} \frac{x}{(1+x^2)^2} dx,
\end{equation}
is improper but convergent because its integrand vanishes as $|x|^{-3} (\beta > 1)$ asymptotically and then it vanishes as its integrand is 
an odd function.
Similarly $<p^2>$ is improper but convergent because its integrand vanishes as $|x|^{-4} (\beta > 1)$
asymptotically and it can be evaluated  as shown below
\begin{equation}
<p^2>=<p\psi_0|p\psi_0>=\frac {\hbar^2}{\pi} \int_{-\infty}^{\infty} \frac{x^2}{(1+x^2)^3} dx= \frac {\hbar^2}{4\pi} \int_{0}^{\pi/2}  [1-\cos4\theta]~ d\theta=\frac{\hbar^2}{8}.
\end{equation}
In the integral above, we have used the substitution $x=\tan \theta$, finally we get a finite value of $\Delta p$ as 
\begin{equation}
\Delta p= \frac{\hbar}{2\sqrt{2}}. 
\end{equation}
Eventually, Eqs. (12,15) render the uncertainty product as  infinite.

Now let us find out from Eq. (6) the potential giving
rise to such a ground state (10), we find
\begin{equation}
V(x)=\frac{2x^2-1}{(1+x^2)^2}.
\end{equation}
This potential is depicted in Fig. 1(c) and the state (10) is its ground state with eigenvalue $E_0=0$. No state can exist below $E=0$ as it  is a node less (ground state). For energies $E>0$ there are scattering states, there may also exist a metastable (quasi bound, resonant state [9]). Hence  the state (10) is the only bound state of this potential (16).

The examples of the ground states $\psi_0(x)$ rendering a finite uncertainty products presented here are instructive which bring out the crucial role of faster asymptotic convergence of $\psi_0(x)$ than $|x|^{-3/2}$. On the other hand the example (10), demonstrates our point that if an eigenstate of a bound state does not vanish faster than $|x|^{-3/2}$, despite being continuous, differentiable and normalizable it would yield the uncertainty product $U$ as infinite.

Similarly, for 3 dimensional central potentials [1,2] in the radial Schr{\"o}dinger equation, where $r \in (0,\infty)$ and $\psi(r)=\frac{u(r)}{r}$, $u(r)$ needs to vanish faster than $r^{-3/2}$ for the uncertainty $\Delta r$ to be finite.

As such  $\Delta x$ or the uncertainty  product becoming infinite does not violate the uncertainty relation (4), this however may not be desirable either. Notice that this ground state (10) is a much extended state that vanishes as $1/x$ asymptotically much slower as compared to the states in Fig. 1(a,b). Such a state, if useful in future,  may be called a loosely bound and spatially extended state.

\end{document}